\documentclass{appolb}
\usepackage{epsfig}

\begin{document}
\pagestyle{plain}
\newcount\eLiNe\eLiNe=\inputlineno\advance\eLiNe by -1
\title{ The dependence structure for PARMA models with $\alpha-$stable innovations
}
\author{J. Nowicka-Zagrajek A. Wy{\l}oma{\'n}ska
\address{Institute of Mathematics and Computer Science\\
Wroc{\l}aw University of Technology\\
Wyb.~Wyspia{\'n}skiego 27, 50-370 Wroc{\l}aw,  Poland}}
\maketitle
\begin{abstract}
In this paper we investigate the dependence structure for PARMA models (i.e. ARMA models with periodic coefficients) with symmetric $\alpha-$stable innovations. In this case the covariance function is not defined and therefore other measures of dependence have to be used. After obtaining the form of the bounded solution of the PARMA system with symmetric $\alpha$-stable innovations, we study the codifference and the covariation -- the most popular measures of dependence defined for symmetric stable time series. We show that both considered measures are periodic. Moreover we determine the cases when the codifference and the covariation are asymptotically proportional with the coefficient of proportionality equal to $\alpha$. 
\end{abstract}

PACS:  89.65.Gh, 05.45.Tp, 02.50.Cw 

\section{Introduction}
Conventional time series analysis is heavily dependent on the assumption of stationarity. But this assumption is unsatisfactory for modelling many real-life phenomena that exhibit seasonal behaviour. Seasonal variations in the mean of time series data can be easily removed by a variety of methods, but when the variance varies with the season, the use of periodic time series models is suggested. In order to model periodicity in autocorrelations, a class of Periodic Autoregressive Moving Average (PARMA) models was introduced. The PARMA$(p,q)$ system is given by:
\begin{eqnarray}\label{parma}
X_n-\sum\limits_{j=1}^pb_j(n)X_{n-j}=\sum_{i=0}^{q-1}a_i(n)\xi_{n-i},
\end{eqnarray}
where $n\in Z$ and the coefficients $\{b_j(n)\}_{j=1}^{p}$ and $\{a_i(n)\}_{i=0}^{q-1}$ are nonzero sequences periodic in $n$ with the same period $T$ while the innovations $\{\xi_n\}$ are independent Gaussian random variables. 
For such PARMA models, the covariance function is a tool for describing the dependence structure of the time series and we can recall that the sequence given by (\ref{parma}) is periodically correlated, more precisely the covariance function $Cov(X_n,X_{n+k})$ is $T-$periodic in $n $ for every $k$ (see \cite{gladyshev, makms91, ww}).
As the coefficients in (\ref{parma}) are periodic, it is obvious that the class of PARMA models is an extension of the class of commonly used ARMA models. 
Due to their interesting properties, PARMA systems have received much attention in the literature and turned out to be an alternative to the conventional stationary time series as they allow for modelling many phenomena in various areas, e.g., in  hydrology (\cite{vec1}), meteorology (\cite{blo}), economics (\cite{bmww, parpag}) and electrical engineering (\cite{garfra}).

The assumption of normality for the observations seems not to be reasonable in the number of applications, such as signal processing, telecommunications, finance, physics and chemistry, and heavy-tailed distributions seem to be more appropriate, see e.g. \cite{mit}. An important class of distributions in this context is the class of $\alpha$-stable (stable) distributions because it is flexible for data modelling and includes the Gaussian distribution as a special case. The importance of this class of distributions is strongly supported by the limit theorems which indicate that the stable distribution is the only possible limiting distribution for the normed sum of independent and identically distributed random variables. Stable random variables have found many practical applications, for instance in finance (\cite{mit}), physics (\cite{janwer}), electrical engineering (\cite{stuckkleiner}).

PARMA models with symmetric stable innovations combine the advantages of classical PARMA models and stable distributions -- they offer an alternative for modelling periodic time series with heavy tails. However, in this case the covariance function is not defined and thus other measures of dependence have to be used. The most popular are the covariation and codifference presented in \cite{st, nw, n}.  

In this paper we consider a special case of stable PARMA models, i.e. PARMA(1,1) systems with symmetric $\alpha$-stable innovations. In this case we rewrite the equation (\ref{parma}) in a simple way, i.e.:
\begin{eqnarray}\label{parma12}
X_n-b_nX_{n-1}=a_n\xi_{n},
\end{eqnarray}
where $n\in Z$ and the coefficients $\{b_n\}$ and $\{a_n\}$ are nonzero  periodic sequences with the same period $T$ while the innovations are independent symmetric $\alpha$-stable (S$\alpha$S for short) random variables given by the following characteristic function:
\begin{eqnarray}\label{innovations}
E\exp(i\theta \xi_n)=\exp(-\sigma^{\alpha}|\theta|^{\alpha}), ~0<\alpha\leq2,
\end{eqnarray}
where $\sigma$ denotes the scale parameter. Let us define $P=b_1 b_2\dots b_T$ and $B_r^s=\prod_{j=r}^sb_j$, with the convention $B_r^s=1$ if $r>s$.

As the covariance function is not defined for stable random vectors, in Section \ref{s_mofdep} we present two other measures of dependence that can be used for symmetric stable time series -- the covariation and the codiffrence. In Section \ref{s_bsol} we discuss the necessary conditions for existence of the bounded solution of PARMA(1,1) systems for $1<\alpha\leq2$ and we note that results obtained in \cite{ww} for PARMA systems with Gaussian innovations can be extended to the case of stable innovations. The covariation and the codifference for PARMA(1,1) models with stable innovations are studied in Section \ref{s_depstr} and the asymptotic relation between these two measures of dependence for the considered models is examined there. We find it interesting to illustrate theoretical results and thus in Section \ref{s_ex} we give an example of PARMA(1,1) systems with S$\alpha$S innovations and illustrate the periodicity of the considered measures of dependence and the asymptotic relation between them.
\section{Measures of dependence for stable time series}\label{s_mofdep}
Let $X$ and $Y$ be jointly S$\alpha$S and let $\Gamma$ be the spectral measure of the random vector $(X,Y)$ (see for instance \cite{st}). If $\alpha<2$ then the covariance is not defined and thus other measures of dependence have to be used. The most popular measures are: the {\bf covariation $CV(X,Y)$} of $X$ on $Y$ defined in the following way:
\begin{eqnarray}
CV(X,Y)=\int_{S_2}s_1s_2^{<\alpha-1>}\Gamma(d{\bf s}), \quad 1<\alpha\leq2,
\end{eqnarray}
where ${\bf s}=(s_1,s_2)$ and the signed power $z^{<p>}$ is given by $z^{<p>}=|z|^{p-1}\bar{z}$, and the {\bf codifference $CD(X,Y)$} of $X$ on $Y$ defined for $0<\alpha\leq2$: 
\begin{eqnarray}
CD(X,Y)=\ln E\exp\{i(X-Y)\}-\ln E\exp\{iX\}-\ln E\exp\{-iY\}.
\end{eqnarray}

Properties of the considered measures of dependence one can find in \cite{st}. Let us only mention here that, in contrast to the codifference, the covariation is not symmetric in its arguments. Moreover, when $\alpha=2$ both measures reduce to the covariance, namely
\begin{eqnarray}\label{covcvcd}
Cov(X,Y)=2CV(X,Y)=CD(X,Y).
\end{eqnarray}

The covariation induces a norm on the linear space of jointly S$\alpha$S random variables $S_{\alpha}$ and this norm is equal to the scale parameter, see \cite{st}. Hence throughout this paper the norm (so-called covariation norm) $||X||_{\alpha}$ is defined by $||X||_{\alpha}=(CV(X,X))^{1/\alpha}$, for a S$\alpha$S random variable $X$ and $\alpha>1$. The sequence $\{X_n\}_{n \in Z}$, is bounded in a space $S_{\alpha}$ with norm $||.||_{\alpha}$ if $\sup_{n \in Z}||X_n||_{\alpha}^{\alpha}<\infty$. Moreover, in this paper we write $X=Y$ in $S_{\alpha}$ if and only if $||X-Y||_{\alpha}=0$. 

If it is possible to transform the sequence $\{X_n\}$ to the moving average representation $X_n=\sum_{j=-\infty}^{\infty}c_j(n)\xi_{n-j}$, where the innovations $\{\xi_n\}$ are independent S$\alpha$S random variables with parameters $\sigma$ and $1<\alpha \leq 2$, then both the covariation and the codifference can be expressed in terms of the coefficients $c_j(n)$  (see \cite{nwylom}): 
\begin{eqnarray*}\label{lemat11}
CV(X_n, X_m)=\sigma^{\alpha}\sum_{j=-\infty}^{\infty}c_j(n)c_{m-n+j}(m) ^{<\alpha-1>},
\end{eqnarray*}
\begin{eqnarray*}\label{lemat12}
CD(X_n, X_m)=\sigma^{\alpha}\sum_{j=-\infty}^{\infty}\left(|c_j(n)|^{\alpha}+|c_{m-n+j}(m)| ^{\alpha}-|c_j(n)-c_{m-n+j}(m)|^{\alpha}\right).
\end{eqnarray*}

\section{Bounded solution of stable PARMA(1,1) system}\label{s_bsol}
In this section we consider PARMA(1,1) system given by (\ref{parma12}) with the S$\alpha$S innovations for $1<\alpha<2$ and we investigate when the bounded solution exists.

Let us assume that $|P|<1$. In this case we show that the considered PARMA(1,1) system has a bounded solution given by the formula:
\begin{eqnarray}\label{forma1}
X_n=\sum\limits_{s=0}^{\infty}B^{n}_{n-s+1}a_{n-s}\xi_{n-s}.
\end{eqnarray}

Provided that every $s\in Z$ can be represented as $s=NT+j$ for some $N=0,1,\dots$ and $j=0,1,\dots, T-1$, we have
\[||X_n||^{\alpha}_{\alpha}=\sigma^{\alpha}\sum\limits_{s=0}^{\infty}\left|B^{n}_{n-s+1}a_{n-s}\right|^{\alpha}=
\sigma^{\alpha}\sum\limits_{N=0}^{\infty}\sum\limits_{j=0}^{T-1}\left|B^{n}_{n-NT-j+1}a_{n-NT-j}\right|^{\alpha}.\]
Now it sufficies to notice that by periodicity of coefficients \mbox{$a_{n-NT-j}=a_{n-j}$} and $B^{n}_{n-NT-j+1}=P^NB^{n}_{n-j+1}$ and that $\sum\limits_{N=0}^{\infty}|P|^{N\alpha}=1/(1-|P|^{\alpha})$. Thus
\[||X_n||^{\alpha}_{\alpha}=\frac{\sigma^{\alpha}}{1-|P|^{\alpha}}\sum\limits_{j=0}^{T-1}\left|B^{n}_{n-j+1}a_{n-j}\right|^{\alpha}
\leq \frac{\sigma^{\alpha}K}{1-|P|^{\alpha}}<\infty,\]
where $K$ is a real constant, $K=\max_{s=1,\dots,T}\{\sum_{j=0}^{T-1}\left|B^{s}_{s-j+1}a_{s-j}\right|^{\alpha}\}$. This implies that $\sup_{n\in Z}||X_n||^{\alpha}_{\alpha}<\infty$ and hence $\{X_n\}$ given by (\ref{forma1}) is bounded. Moreover, it is easy to check that $\{X_n\}$ given by (\ref{forma1}) satisfies equation (\ref{parma12}). 

To prove the converse, let us note that iterating the equation (\ref{parma12}) yields:
\[X_{n}=B^{n}_{n-k+1}X_{n-k}+\sum_{s=0}^{k-1}B^{n}_{n-s+1}a_{n-s}\xi_{n-s}.\]
For each $n\in Z$ and $k=NT+j$ by periodicity of the coefficients we have $B_{n-k+1}^n=P^N B_{n-j+1}^n$. Therefore, if $|P|<1$ and $\{X_n\}$ is a bounded solution of (\ref{parma12}) then 
\begin{eqnarray*}\label{lim}
\lim_{k\rightarrow\infty}||X_n-\sum_{s=0}^{k-1}B_{n-s+1}^{n}a_{n-s}\xi_{n-s}||_{\alpha}=\lim_{k\rightarrow\infty}||X_{n-k}B^{n}_{n-k+1}||_{\alpha}=0.
\end{eqnarray*}
This means that the bounded solution of the system considered in this section is given by
\[X_n= \lim_{k\rightarrow\infty}\sum_{s=0}^{k-1}B^{n}_{n-s+1}a_{n-s}\xi_{n-s}=\sum_{s=0}^{\infty}B_{n-s+1}^{n}a_{n-s}\xi_{n-s}.\]

In a similar manner, it can be shown, that if $|P|>1$, then the bounded solution of the system under study is given by
\begin{eqnarray}\label{forma2}
X_n=-\sum\limits_{s=1}^{\infty}\frac{a_{n+s}}{B^{n+s}_{n+1}}\xi_{n+s}.
\end{eqnarray}

It is worth pointing out here that the solution for the PARMA(1,1) system with S$\alpha$S innovations takes the same form of the moving average as in the case of Gaussian innovations obtained in \cite{ww} and it reduces to well-known formula for ARMA models in case of constant coefficients. 
\section{Dependence structure of stable PARMA(1,1) models}\label{s_depstr}
Let us consider PARMA(1,1) system given by (\ref{parma12}) with S$\alpha$S innovations for $1<\alpha\leq2$.  Moreover, we will restrict our attention to the case $|P|<1$ because this case is more important for applications -- as shown in Section \ref{s_bsol}, the considered system has a bounded solution $\{X_n\}$ in the linear space of jointly S$\alpha$S random variables with norm $||.||_{\alpha}$ given by the causal moving average representation (\ref{forma1}).

In order to investigate the dependence structure of stable PARMA(1,1) models we will first rewrite $X_n$ as
$X_n=\sum_{s=-\infty}^{\infty}c_s(n)\xi_{n-s}$, where
\begin{eqnarray}\label{cspmn1}
  c_s(n)=
\left\{\begin{array}{ll} 0, & \mbox{if}~ s<0,\\
B^{n}_{n-s+1}a_{n-s},& \mbox{if}~ s\geq 0,
\end{array}\right.
\end{eqnarray}
and then using results presented in Section \ref{s_mofdep} we will find formulas for the covariation and the codifference. We will also study periodicity of these measures and their asymptotic relation.

{\bf The covariation.}
If $n\geq m$, then for $1<\alpha \leq 2$ we have
\[CV(X_n,X_{m})=\sigma^{\alpha}\sum\limits_{j=n-m}^{\infty}B^n_{n-j+1}a_{n-j}\left(B^{m}_{n-j+1}a_{n-j}\right)^{<\alpha-1>}.\]
It is easy to notice that $z\cdot z^{<\alpha-1>}=|z|^{\alpha}$ and $B^{n}_{n-j+1}=B^{n}_{m+1}B^{m}_{n-j+1}$. Therefore, for $s=j-n+m$ we obtain
\[CV(X_n,X_{m})=\sigma^{\alpha}B_{m+1}^{n}\sum\limits_{s=0}^{\infty}\left|B_{m-s+1}^{m}a_{m-s}\right|^{\alpha}.\]
As every $s\in Z$ can be represented as $s=NT+j$ for some $N=0,1,\dots$ and $j=0,1,\dots, T-1$, we have
\[CV(X_n,X_{m})=\sigma^{\alpha}B_{m+1}^{n}\sum\limits_{N=0}^{\infty}\sum\limits_{j=0}^{T-1}\left|B_{m-NT-j+1}^{m}a_{m-NT-j}\right|^{\alpha}.\]
Now it suffices to notice that by periodicity of coefficients \mbox{$a_{m-NT-j}=a_{m-j}$} and $B^{m}_{m-NT-j+1}=P^NB^{m}_{m-j+1}$ and that $\sum\limits_{N=0}^{\infty}|P|^{N\alpha}=1/(1-|P|^{\alpha})$. Thus
\begin{eqnarray}\label{cv1}
CV(X_n,X_{m})=\sigma^{\alpha}\frac{B_{m+1}^{n}}{1-|P|^{\alpha}}\sum\limits_{j=0}^{T-1}\left|B_{m-j+1}^{m}a_{m-j}\right|^{\alpha}.
\end{eqnarray}
If $n<m$, then the covariation is given by
\[CV(X_n,X_{m})=\sigma^{\alpha}\sum\limits_{j=0}^{\infty}B^{n}_{n-j+1}a_{n-j}\left(B^{m}_{n-j+1}a_{n-j}\right)^{<\alpha-1>}.\]
In this case $B^{n}_{n-j+1}=B^{m}_{n-j+1}/B^{m}_{n+1}$ which results in formula
\[CV(X_n,X_{m})=\frac{\sigma^{\alpha}}{B_{n+1}^{m}}\sum\limits_{j=0}^{\infty}\left|B_{n-j+1}^{m}a_{n-j}\right|^{\alpha}.\]
The simple calculation similar like in the previous case leads us to the result
\begin{eqnarray}\label{cv2}
CV(X_n,X_{m})=\frac{\sigma^{\alpha}}{B_{n+1}^{m}(1-|P|^{\alpha})}\sum\limits_{j=0}^{T-1}\left|B_{n-j+1}^{m}a_{n-j}\right|^{\alpha}.
\end{eqnarray}
As the sequences $\{a_n\}$ and $\{b_n\}$ are periodic in $n$ with period $T$, the covariation is periodic in $n$ and $m$ with the same period, indeed $CV(X_n,X_m)=CV(X_{n+T},X_{m+T})$.

{\bf The codifference.}
For $1<\alpha \leq 2$ and $n\geq m$ we have to calculate
\[\sum\limits_{j=n-m}^{\infty}\left(\left|B^n_{n-j+1}a_{n-j}\right|^{\alpha}+\left|B^{m}_{n-j+1}a_{n-j}\right|^{\alpha}-\left|B^n_{n-j+1}a_{n-j}-B^m_{n-j+1}a_{n-j}\right|^{\alpha}\right).\]
Observe that there are $a_{n-j}$ and $B^{m}_{n-j+1}$ (as $B^{n}_{n-j+1}=B^{m}_{n-j+1}B^{n}_{m+1}$) in each part of the above formula. Therefore we can write
\[CD(X_n,X_{m})=\sigma^{\alpha}\left(1+|B_{m+1}^n|^{\alpha}-|1-B_{m+1}^n|^{\alpha}\right)\sum\limits_{j=n-m}^{\infty}|B_{n-j+1}^{m}a_{n-j}|^{\alpha},\]
that can be transformed to
\[CD(X_n,X_{m})=\sigma^{\alpha}\left(1+|B_{m+1}^n|^{\alpha}-|1-B_{m+1}^n|^{\alpha}\right)\sum\limits_{s=0}^{\infty}|B_{m-s+1}^{m}a_{m-s}|^{\alpha}.\]
And now it is sufficient to notice that the sum $\sum_{s=0}^{\infty}|B_{m-s+1}^{m}a_{m-s}|^{\alpha}$ has been already calculated for the covariation. So we obtain
\begin{eqnarray}\label{cd1}
CD(X_n,X_{m})=\sigma^{\alpha}\frac{1+|B_{m+1}^n|^{\alpha}-|1-B_{m+1}^n|^{\alpha}}{1-|P|^{\alpha}}\sum\limits_{j=0}^{T-1}|B_{m-j+1}^{m}a_{m-j}|^{\alpha}.
\end{eqnarray}
It is not necessary to calculate the codifference for $n<m$ as this measure is symmetric in its arguments, i.e. for every $n,m\in Z$ we have $CD(X_n,X_{m})=CD(X_m,X_{n})$. 
By the assumption, the coefficients $\{a_n\}$ and $\{b_n\}$ are periodic in $n$ with period $T$. Thus it is not difficult to notice that 
$CD(X_n,X_m)=CD(X_{n+T},X_{m+T})$, which means that the codifference is periodic in $n$ and $m$ with the period $T$.

{\bf Asymptotic relation between CV and CD.}
Using formulas (\ref{cv1}), (\ref{cv2}) and (\ref{cd1}) and two simple facts that hold for $1<\alpha<2$, $0<|P|<1$ and any real constant $a\neq0$:
\[\lim\limits_{m\rightarrow\infty}\frac{1+|aP^m|^{\alpha}-\left|1-aP^m\right|^{\alpha}}{aP^m}=\alpha,\]
\[\lim\limits_{m\rightarrow\infty}\frac{aP^m\left(1+\left|aP^m\right|^{\alpha}-\left|1-aP^m\right|^{\alpha}\right)}{\left|aP^m\right|^{\alpha}}=0,\]
it can be proved that for the considered PARMA(1,1) systems with S$\alpha$S innovations the following relations hold for each $n\in Z$ and $1<\alpha<2$: 
\begin{eqnarray}\label{glowne11}
\lim\limits_{k\rightarrow \infty}\frac{CD(X_n,X_{n-k})}{CV(X_n,X_{n-k})}=\lim\limits_{k\rightarrow \infty}\frac{CD(X_{n+k},X_{n})}{CV(X_{n+k},X_{n})}=\alpha,
\end{eqnarray}
\begin{eqnarray}\label{glowne12}
\lim\limits_{k\rightarrow \infty}\frac{CD(X_{n-k},X_{n})}{CV(X_{n-k},X_{n})}=\lim\limits_{k\rightarrow \infty}\frac{CD(X_{n},X_{n+k})}{CV(X_{n},X_{n+k})}=0.
\end{eqnarray}

The interesting point is that, because of asymmetry of the covariance, we obtained quite different asymptotic results. Moreover, (\ref{glowne11}) extends the results obtained in \cite{nw} for ARMA models and it reduces to (\ref{covcvcd}) for $\alpha=2$.

\section{Example}\label{s_ex}
\begin{figure}[tb]
\begin{center}
\includegraphics[width=9cm]{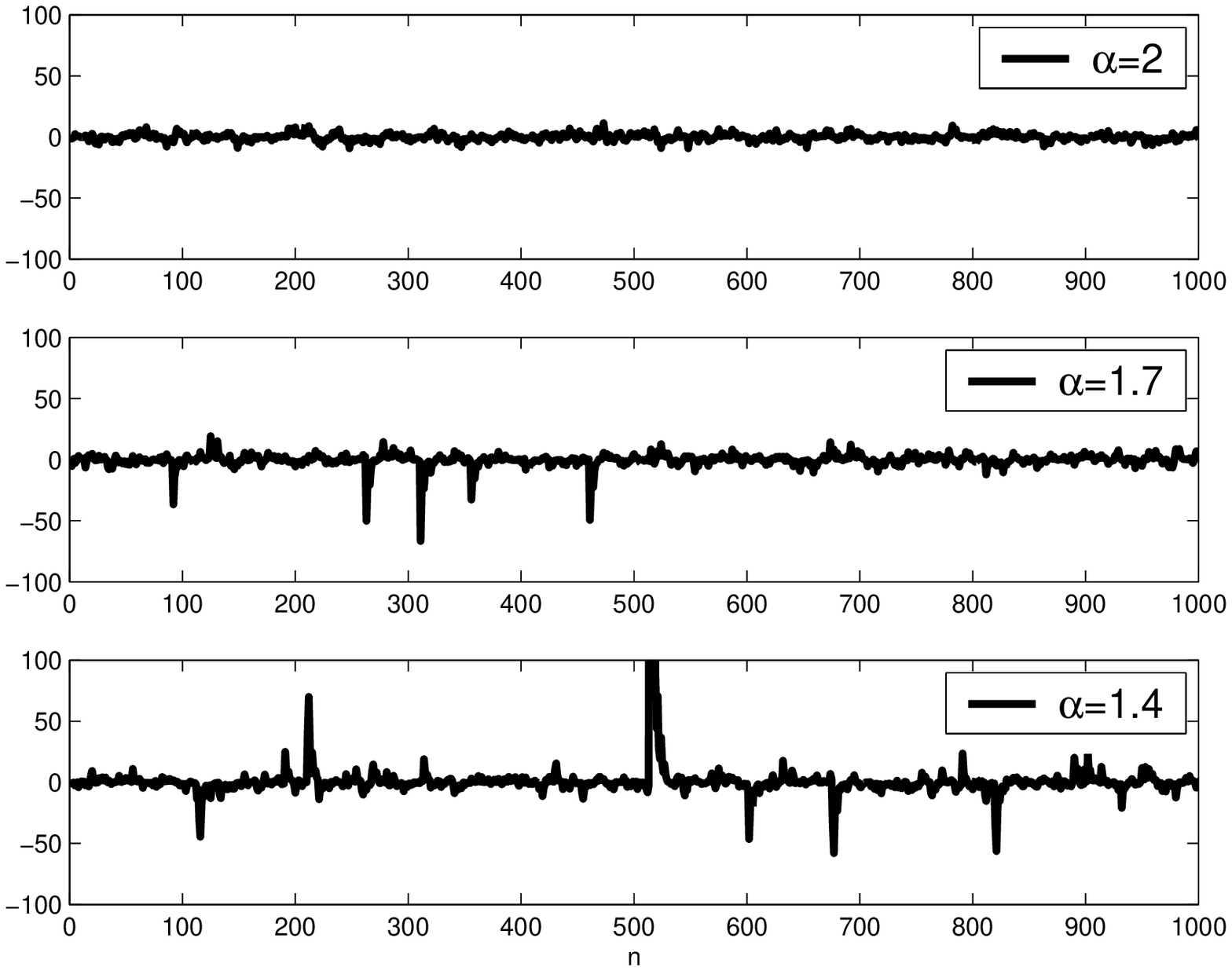} \caption[The realizations of PARMA(1,1) model with S$\alpha$S innovations for $\alpha=2$ (top panel), $\alpha=1.7$ (middle panel) and $\alpha=1.4$ (bottom panel).]{}\label{fig_real}
\end{center}
\end{figure}
In order to illustrate our theoretical results let us consider PARMA(1,1) model with S$\alpha$S innovations with $\sigma=1$, where the coefficients are given by  
\begin{eqnarray*}
b_n = \left\{
\begin{array}{ll}  \displaystyle
  0.5 &\mbox{if} ~ n=1,4,7,\dots,\\
\displaystyle 1.6,& \mbox{if} ~ n=2,5,8,\dots,\\
\displaystyle 0.4,& \mbox{if} ~ n=3,6,9,\dots,\\
\end{array}
\right. \quad a_n = \left\{
\begin{array}{ll}  \displaystyle
  1& \mbox{if} ~ n=1,4,7,\dots,\\
\displaystyle 2,& \mbox{if} ~ n=2,5,8,\dots,\\
\displaystyle 0.003,& \mbox{if} ~ n=3,6,9,\dots.\\
\end{array}
\right.
\end{eqnarray*}

\begin{figure}[tb]
\begin{center}
\includegraphics[width=9cm]{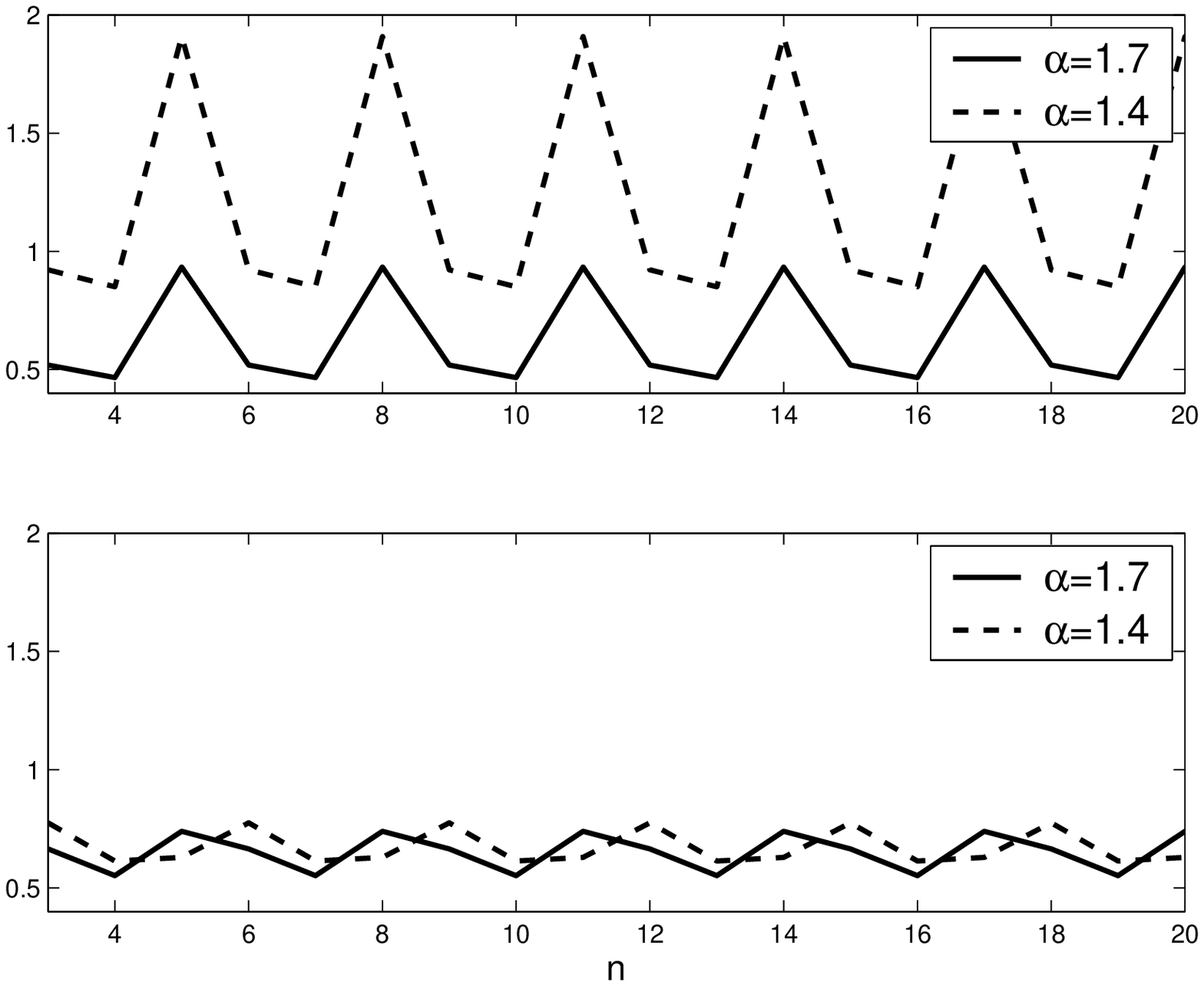} \caption[The codifference $CD(X_n,X_{n+k})$ (top panel) and the covariation $CV(X_n,X_{n+k})$ (bottom panel) for PARMA(1,1) model with S$\alpha$S innovations for $\alpha=1.7$ (solid line) and $\alpha=1.4$ (dotted line) vs $ n=3,4,\dots,20$ for $k=5$.]{}\label{fig_miary}
\end{center}
\end{figure}
It is clear that the coefficients are periodic with $T=3$ and in this case $P=0.32$. Therefore we are allowed to use formulas obtained in Sections \ref{s_bsol} and \ref{s_depstr}. We first want to demonstrate how the parameter $\alpha$ influences the behaviour of the time series, so  we plot 1000 realizations of the considered model for $\alpha=2$, $\alpha=1.7$ and $\alpha=1.4$, see Figure \ref{fig_real}. It is easy to notice that the smaller $\alpha$ we take, the greater values of the time series can appear (property of heavy-tailed distributions). Next, we want to show the dependence structure of the considered model, especially periodicity of measures of dependence. In order to do this we plot the codifference $CD(X_n,X_{n+k})$ and the covariation $CV(X_n,X_{n+k})$ for $\alpha=1.7$ and $\alpha=1.4$ in case of $k=5$, see Figure \ref{fig_miary}. Although the behaviour of the measures of dependence depends on the parameter $\alpha$, one can observe that both measures are periodic with the same period $T=3$.

\begin{figure}[tb]
\begin{center}
\includegraphics[width=9cm]{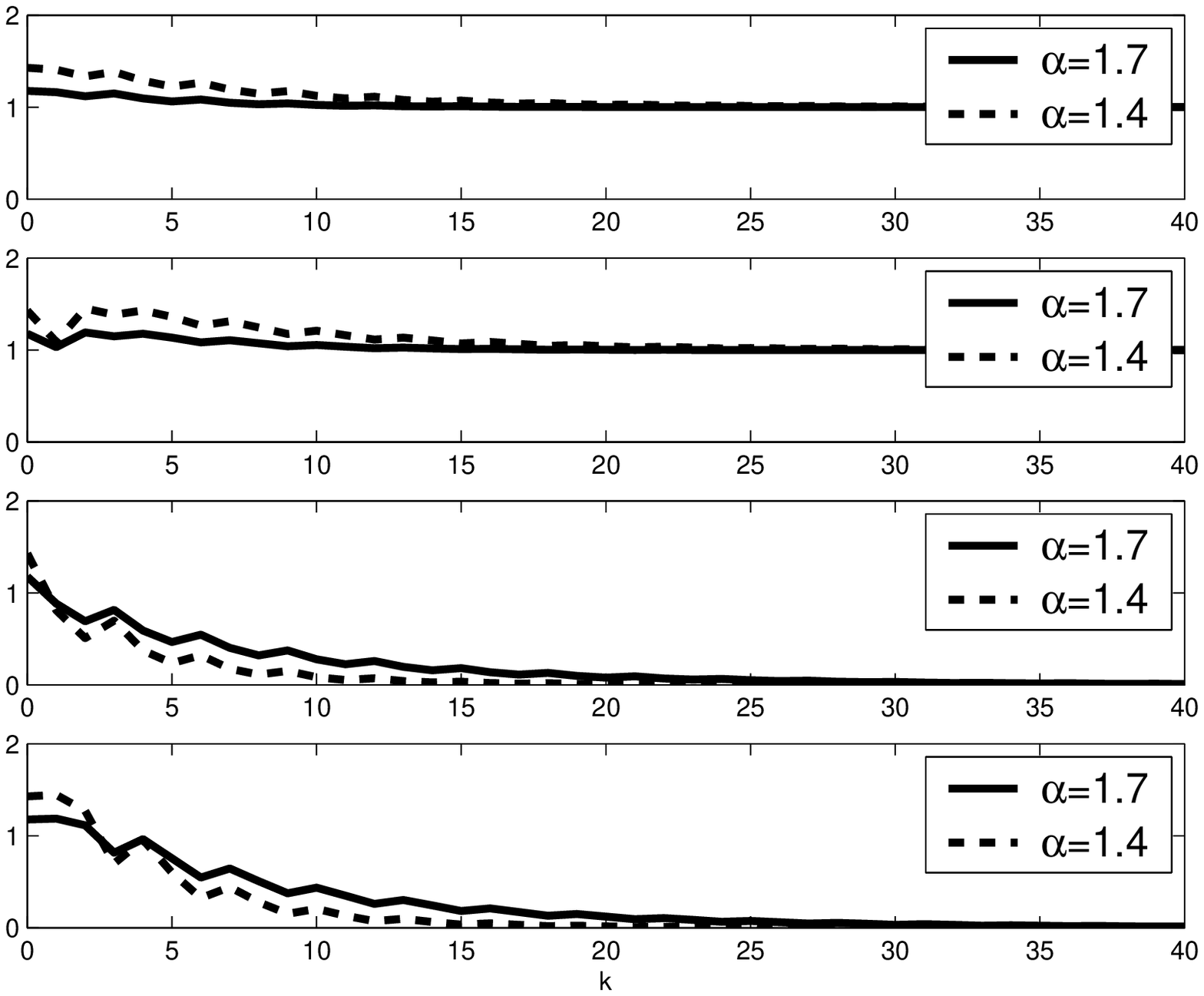} \caption[The plots of the functions $\frac{CD(X_{n+k},X_{n})}{\alpha CV(X_{n+k},X_{n})}$ (the first panel), $\frac{CD(X_{n},X_{n-k})}{\alpha CV(X_{n},X_{n-k})}$ (the second panel), $\frac{CD(X_{n},X_{n+k})}{\alpha CV(X_{n},X_{n+k})}$ (the third panel) and $\frac{CD(X_{n-k},X_{n})}{\alpha CV(X_{n-k},X_{n})}$ (the fourth panel) vs $ k=0,1,\dots,40 $ for $n=50$, $\alpha=1.7$ (solid line) and $\alpha=1.4$ (dotted line).]{}\label{fig_asym}
\end{center}
\end{figure}

Finally, let us illustrate the asymptotic relation between the covariation and the codifference that is studied in Section \ref{s_depstr}. Figure \ref{fig_asym} contains plots of the functions $\frac{CD(X_{n+k},X_{n})}{\alpha CV(X_{n+k},X_{n})}$, $\frac{CD(X_{n},X_{n-k})}{\alpha CV(X_{n},X_{n-k})}$, $\frac{CD(X_{n},X_{n+k})}{\alpha CV(X_{n},X_{n+k})}$ and $\frac{CD(X_{n-k},X_{n})}{\alpha CV(X_{n-k},X_{n})}$ for $ k=0,1,\dots,40 $, $n=50$ and $\alpha=1.7$ and $\alpha=1.4$. 
According to the theoretical results, the first two quotients tend to 1 and the next two tend to 0 as $k$ increases.

\listoffigures

\end{document}